\documentclass[reqno]{amsproc}

\usepackage{bm,amsfonts, mathtools}
\usepackage{graphicx,color, wasysym}
\usepackage{textcomp}
\usepackage{amsmath,amssymb,latexsym,epsfig}
\usepackage[british]{babel}
\usepackage{hyphenat}
\usepackage{appendix}
\usepackage{txfonts}

\def\ulamek#1#2{\mbox{\normalfont$\frac{#1}{#2}$}}

\DeclareMathOperator{\D}{d\!}
\DeclareMathOperator{\E}{e} 
\DeclareMathOperator{\I}{i}
   \DeclareMathOperator{\RE}{\mathfrak{Re}}

\begin{document} 



\title{On the complete monotonicity of the three parameter generalized Mittag-Leffler function 
$E_{\alpha, \beta}^{\gamma}(-x)$}

\author{K. G\'{o}rska}
\address{H. Niewodnicza\'{n}ski Institute of Nuclear Physics, Polish Academy of Sciences, Division of Theoretical Physics, ul. Eliasza-Radzikowskiego 152, PL 31-342 Krak\'{o}w, Poland}
\email{katarzyna.gorska@ifj.edu.pl}

\author{A. Horzela}
\address{H. Niewodnicza\'{n}ski Institute of Nuclear Physics, Polish Academy of Sciences, Division of Theoretical Physics, ul. Eliasza-Radzikowskiego 152, PL 31-342 Krak\'{o}w, Poland}
\email{andrzej.horzela@ifj.edu.pl}

\author{A. Lattanzi}
\address{H. Niewodnicza\'{n}ski Institute of Nuclear Physics, Polish Academy of Sciences, Division of Theoretical Physics, ul. Eliasza-Radzikowskiego 152, PL 31-342 Krak\'{o}w, Poland}
\email{ambra.lattanzi@ifj.edu.pl}

\author{T. K. Pog\'{a}ny}
\address{Faculty of Maritime Studies, University of Rijeka, Studentska 2, HR-51000 Rijeka, Croatia\\
Institute of Applied Mathematics, \'Obuda University, B\'ecsi \'ut 96/b, H-1034 Budapest, Hungary}
\email{poganj@pfri.hr}

\begin{abstract}
Using the Bernstein theorem we give a simple proof of the complete monotonicity of the three parameter generalized Mittag-Leffler function $E_{\alpha, \beta}^{\gamma}(-x)$ for $x \geq 0$ and suitably adjusted parameters $\alpha$, $\beta$ and $\gamma$.
\end{abstract}

\maketitle


The three parameter generalization of the Mittag-Leffer function, in the literature bearing the name of the Prabhakar 
function \footnote{In physically oriented papers this name is sometimes used also for $x^{\beta-1}E_{\alpha, \beta}^{\gamma}(-x^{\alpha})$.}, was introduced by T. R. Prabhakar \cite{Prabhakar} as 
\begin{equation}\label{1}
E_{\alpha, \beta}^{\gamma}(z) = \sum_{r=0}^{\infty} \frac{(\gamma)_{r}}{r! \Gamma(\alpha r + \beta)} z^{r}
\end{equation}
where $z\in\mathbb{C}$ , $\RE(\alpha) > 0$ and the Pochhammer symbol $(\gamma)_{r}$ denotes $\Gamma(\gamma + r)/\Gamma(\gamma)$. For $\beta=\gamma=1$ it reduces to the standard Mittag-Leffler function $E_{\alpha}(z)$, while for $\gamma=1$ becomes two parameter generalized Mittag-Leffler function  $E_{\alpha, \beta}(z)$, named also the Wiman one \cite{book1}. Our aim is to give a simple proof of the complete monotonicity of $E_{\alpha, \beta}^{\gamma}(-x)$ for $x\ge 0$ and suitably adjusted parameters $\alpha, \beta$  and $\gamma$.

Properties of functions belonging the Mittag-Leffler family, especially those involving complete monotonicity (for the first time being studied in classical papers by H. Pollard \cite{Pollard48} and W. R. Schneider  \cite{Schneider96})  are interesting both for mathematicians and physicists. For the first of these communities various Mittag-Leffler  functions are important objects in the theory and applications of special functions \cite{Samko,Shukla,Simon}, fractional calculus and fractional differential equations \cite{book1, Mainardibook}, while the representatives of the second community successfully use Mittag-Leffler functions to describe phenomena exhibiting the memory-dependent time evolution, like dielectric relaxation \cite{Hanyga1, Mainardi16} and/or viscoelasticity \cite{Mainardibook,Hanyga2}. The complete monotonicity enables using the Bernstein theorem (see, e.g., \cite{Widder} Theorem 12a, \cite{Schilling} Theorem 1.4) which leads to one to one correspondence between completely monotonic functions and Laplace transforms of probability measures supported on $\mathbb{R}_{+}$. This has a straightforward physical interpretation - time decay patterns, if given by  completely monotone functions of time, may be represented as effects of summing up elementary Debye (exponential)  decays $\exp(-t/\tau)$ weighted by distributions of characteristic decay times $\tau$. An example of such approach is  provided by a commonly met relaxation pattern called the Havriliak-Negami model \cite{Havriliak}. The latter is based on the phenomenologically observed spectral function in the (complex) frequency domain  $(1 + s^{\alpha})^{-\gamma}$, where $s=i\omega\tau^{*}$ with $\omega$ being the frequency and $\tau^{*}$ denotes some effective characteristic time \cite{{CJFBottcher78}}. Taking the inverse Laplace transform of the formula (2.5) in \cite{Prabhakar}
\begin{equation}\label{2}
\mathcal{L}[t^{\beta - 1} E^{\gamma}_{\alpha, \beta}(- x t^{\alpha})](s) = \frac{s^{\alpha\gamma-\beta}}{(x + s^{\alpha})^{\gamma}},
\end{equation}
valid for $\RE(s), \RE(\beta) > 0$ and $|s| > |x|^{1/\RE(\alpha)}$, we conclude that  the three parameter generalized Mittag-Leffler function $E^{\gamma}_{\alpha, \alpha\gamma}(- x t^{\alpha})$, taken for $\beta = \alpha\gamma$ and multiplied by a prefactor $t^{\alpha\gamma - 1}$, may be interpreted as a response function giving the time decay of polarization induced in some dielectric medium described using the Havriliak-Negami pattern  \cite{Mainardibook,KGorska18}. This property links the Mittag-Leffler functions and the Havriliak-Negami  model and provokes to ask the physically motivated question:  {\em Do such obtained functions result from summing up the Debye relaxation  processes?}  To answer it, in view of the Bernstein theorem, is equivalent to judging the problem of complete monotonicity of the function $t^{\alpha\gamma - 1}E^{\gamma}_{\alpha, \alpha\gamma}(- x t^{\alpha})$.

Properties of the function $t^{\beta - 1} E^{\gamma}_{\alpha, \beta}(- x t^{\alpha})$ have been recently the subject of extensive investigations \cite{Capelas,Tomovski,Mainardi15} where the authors have shown that $t^{\beta - 1} E^{\gamma}_{\alpha, \beta}(- x t^{\alpha})$ is completely monotone for $x,t\ge 0$ and parameters satisfying $0<\alpha, \beta, \gamma \le 1$  and $0 < \alpha\gamma\le \beta \le 1$. Proving the complete monotonicity of $t^{\beta - 1} E^{\gamma}_{\alpha, \beta}(- x t^{\alpha})$ the authors of \cite{Capelas,Tomovski,Mainardi15} have treated this function as a whole and have not investigated the complete monotonicity of the function $E^{\gamma}_{\alpha, \beta}(- x )$ itself, leaving a gap between their results and the classical ones of Pollard \cite{Pollard48} and Schneider \cite{Schneider96} who, in 1948 and 1996 respectively, demonstrated  the complete monotonicity of the Mittag-Leffler function $E_{\alpha}(- x )$ for $x\ge 0$, $0\le\alpha\le 1$ and of the Wiman function $E_{\alpha, \beta}(- x )$ for  $x\ge 0$, $0\le\alpha\le 1$ and $\beta$ restricted only by the condition $\beta\ge\alpha$. Thus, according to the Schneider result, the complete monotonicity of $E_{\alpha, \beta}(- x )$ does not require any upper bound put on the parameter $\beta$. This observation has motivated us to investigate the complete monotonicity of $E^{\gamma}_{\alpha, \beta}(- x )$ and to clarify the meaning of the conditions $0\le\beta\le 1$ and $0 < \alpha\gamma\le \beta \le 1$ obtained in \cite{Capelas,Tomovski,Mainardi15} to guarantee the complete monotonicity of $t^{\beta - 1} E^{\gamma}_{\alpha, \beta}(- x t^{\alpha})$.

Inverting the Laplace transform Eq. \eqref{2}  and extracting the Mittag-Leffler function we have
   \begin{equation}\label{3}
      E^{\gamma}_{\alpha, \beta}(- xt^{\alpha}) = \frac{t^{1-\beta}}{2\pi\!\I}\int_{L_s} \E^{st} 
			   \frac{s^{\alpha\gamma-\beta}}{(s^{\alpha} + x)^{\gamma}} \D s
   \end{equation}
where $L_s$ is the Bromwich contour with $\RE(s) > 0$. Changing the variable $st = \xi^{1/\alpha}$ and next using formula 2.3.3.1 of \cite{Prudnikov1}  
   \begin{equation} \label{FX}
	    (\xi + xt^{\alpha})^{-\gamma} = \frac1{\Gamma(\gamma)}\int_{0}^{\infty} 
			                                \E^{-(\xi + xt^{\alpha}) u} u^{\gamma-1} \D u\,,
   \end{equation}
we arrive at
   \begin{equation}\label{4}
      E^{\gamma}_{\alpha, \beta}(- xt^{\alpha}) = \frac{1}{2\pi\!\I\alpha}\int_{L_\xi} \E^{\xi^{1/\alpha}} 
			   \xi^{\frac{\alpha\gamma - \beta + 1}{\alpha} - 1} \left[\frac{1}{\Gamma(\gamma)}\int_{0}^{\infty} 
				 \E^{-(\xi + xt^{\alpha}) u} u^{\gamma-1} \D u\right]\D\xi.
   \end{equation}
These steps are legitimate as by the substitution $\xi = (st)^\alpha$ one transforms the integration contour $L_s$ into $L_\xi$ and obviously $\RE(\xi) = |s|^\alpha t^\alpha \cos\big(\alpha \arg(s)\big)>0$, therefore the convergence of integral \eqref{FX} is controlled. Both integrals in Eq. \eqref{4} converge absolutely so we are allowed to change the order of in\-teg\-ration. 
This gives
   \begin{equation}\label{5}
      E^{\gamma}_{\alpha, \beta}(-xt^{\alpha}) = \frac{1}{\alpha} \int_{0}^{\infty} \E^{- x t^{\alpha}u}  
	       u^{-1-\frac{1}{\alpha}}\, g^{\gamma}_{\alpha, \beta}(u^{-1/\alpha}) \D u, 
   \end{equation}
{with}
  \begin{equation} \label{51}
      g^{\gamma}_{\alpha, \beta}(y) = \frac{y^{-1-\alpha\gamma}}{2\pi\!\I \Gamma(\gamma)} 
	                                  \int_{L_\xi} \E^{\xi^{1/\alpha}} \E^{- \xi y^{-\alpha}} 
\xi^{\frac{\alpha\gamma-\beta + 1}{\alpha}-1} \D\xi.
   \end{equation}
Let's put  $\xi y^{-\alpha} = z^{\alpha}$ in the definition of $g^{\gamma}_{\alpha, \beta}(y)$. Then
   \begin{equation}\label{6}
      g^{\gamma}_{\alpha, \beta}(y) = \frac{\alpha}{\Gamma(\gamma)} y^{-\beta} f_{\alpha, \beta}^{\gamma}(y),  
				 \qquad f_{\alpha, \beta}^{\gamma}(y) = \frac1{2\pi\!\I}\,\int_{L_z} \E^{y z} z^{\alpha\gamma-\beta} 
				 \E^{-z^{\alpha}} \D z,
   \end{equation}
where the function $f_{\alpha, \beta}^{\gamma}(y)$ is an auxiliary quantity which we will use to prove the nonnegativity of $g_{\alpha, \beta}^{\gamma}(y)$ for $y > 0$. We remind, recalling once more the Bernstein theorem, that $f_{\alpha, \beta}^{\gamma}(y)$ would be nonnegative on $\mathbb{R}_{+}$ iff its Laplace transform is completely monotone.  The latter, because of the second expression in Eq. \eqref{6}, reads 
   \begin{equation}\label{7}
      \int_{0}^{\infty}  \E^{-y z} f_{\alpha, \beta}^{\gamma}(y) \D y = z^{\alpha\gamma-\beta} \E^{-z^{\alpha}}.
   \end{equation}
The general property of completely monotone functions says that the product of two completely monotone functions also obeys  this property. So it  is enough to show that $\E^{-z^{\alpha}}$ and $z^{\alpha\gamma-\beta}$ are completely monotone. The complete monotonicity of the stretched exponential $\E^{-z^{\alpha}}$ for $0 < \alpha < 1$ was  shown in \cite{Pollard46}, whereas $z^{\alpha\gamma-\beta}$ is completely monotone if  $\gamma > 0$ and $\beta\geq \alpha\gamma$. Thus for such parameters $f_{\alpha, \beta}^{\gamma}(y)$ is nonnegative on $\mathbb{R}_{+}$ and so is the function $u^{\frac{1}{\alpha}(\beta - \alpha -1)}f_{\alpha, \beta}^{\gamma}(u^{-\frac{1}{\alpha}})$ in Eq. \eqref{5};  renaming there $xt^{\alpha}$ as $w$ we see that $E^{\gamma}_{\alpha, \beta}(-w)$ is the Laplace transform of a nonnegative function which ends up the proof of its complete monotonicity.
      
The obtained results enable us to claim that the three parameter generalized Mittag-Leffler function $E_{\alpha, \beta}^{\gamma}(-x)$ is completely monotone if  $0 < \alpha < 1$, $\gamma > 0$, and $\beta \geq \alpha\gamma$. We point out that the restriction $\beta < 1$ (appearing in \cite{Capelas,Tomovski,Mainardi15}) explains the complete monotonicity of $t^{\beta-1}E_{\alpha, \beta}^{\gamma}(-t^{\alpha})$ as a result of completely monotone substitution  $x \to t^{\alpha}$ in $E_{\alpha, \beta}^{\gamma}(-x)$ \footnote{A substitution $(f\circ g)(x)$ of a completely monotone function $g(x)$ into another completely monotone function $f(x)$ leads to the complete monotonicity of the result. } and next taking the product of completely monotone functions. However it is {\em not needed} for $E_{\alpha, \beta}^{\gamma}(-x)$ itself;  observe also that our conditions for $\alpha, \beta$ and $\gamma$ fully coincide with that previously known as leading to the complete monotonicity of the Mittag-Leffler function $E_{\alpha}(-x)=E_{\alpha,1}^{1}(-x)$ ($0 < \alpha < 1$, \cite{Pollard48}) and of the two parameter generalized Mittag-Leffler function $E_{\alpha, \beta}(-x)=E_{\alpha,\beta}^{1}(-x)$ ($0 < \alpha < 1$ and $\beta\geq\alpha$, \cite{Schneider96}).

The auxiliary function $f_{\alpha, \beta}^{\gamma}(y)$ given by the inverse Laplace transform of Eq. \eqref{6}, if taken for rational $\alpha = l/k < 1$, can be written down (see Eq. (2.2.1.19) of \cite{APPrudnikov-v5}) in terms of the Meijer $G$-function
\begin{equation}\label{8}
f_{l/k, \beta}^{\gamma}(y) = \frac{\sqrt{l k}}{(2\pi)^{\frac{k-l}{2}}} \frac{l^{\frac{l}{k}\gamma - \beta}}{y^{1 + \frac{l}{k}\gamma - \beta}}\; G_{l, k}^{k, 0} \left(\frac{l^{l}}{k^{k} y^{l}}\,\Bigg\vert {\Delta(l, \beta - \frac{l}{k}\gamma) \atop \Delta(k, 0)}\right),
\end{equation}
where the symbol $\Delta(n, a)$ denotes the list of parameters  $a/n, (a+1)/n, \ldots, (a+n-1)/n$.\footnote{Eq. \eqref{8} is listed without a proof as a special case of formula (2.2.1.19) in \cite{APPrudnikov-v5} taken for for $\nu = \beta - \frac{l}{k}\gamma$ and $a = 1$.} For some values of $\alpha$, $\beta$, and $\gamma$ the function $f_{\alpha, \beta}^{\gamma}(y)$'s can be expressed in more familiar ways, e.g.: 
\begin{itemize}
\item[(a)] if $0 < \alpha=l/k < 1$ and $\beta = {\gamma l}/k  > 0$, with no further restrictions put on the parameters $\beta$ or $\gamma$,  the functions $f_{\alpha, \gamma \alpha}^{\gamma}$ are proportional to the one-sided L\'{e}vy stable distributions \cite{KGorska18} which explicit forms may be found in \cite{Pollard46, KAPenson10, KGorska12, GDattoli14} and in references therein, 

\item[(b)] for $\alpha = 1/2$, $\beta = 3/2$, and $\gamma = 2$ one gets $f_{1/2, 3/2}^{2}(y) = \E^{-1/(4y)}/ \sqrt{\pi y}$; positive for $y\geq 0$,

\item[(c)] $f_{\alpha, \beta}^{\gamma}(y)$ may be related to the modified Bessel function of the second kind: $\alpha = 1/3$, $\beta = 2$, and $\gamma = 5$ we have $f_{1/3, 2}^{5}(y) = \frac{1}{\pi y \sqrt{3}} K_{2/3}\!\left(\ulamek{2}{3 \sqrt{3 y}}\right)$, while for  $\alpha = 2/3$, $\beta = 2$, and $\gamma = 2$ one finds $f_{2/3, 2}^{2}(y) = \frac{1}{\sqrt{3} \pi y} \E^{-2/(27 y^{2})} K_{1/3}(\ulamek{2}{27 y^{2}})$; both being positive for $y\geq 0$. 
\end{itemize}

\begin{figure}[!h]
\begin{center}
\includegraphics[scale=0.6]{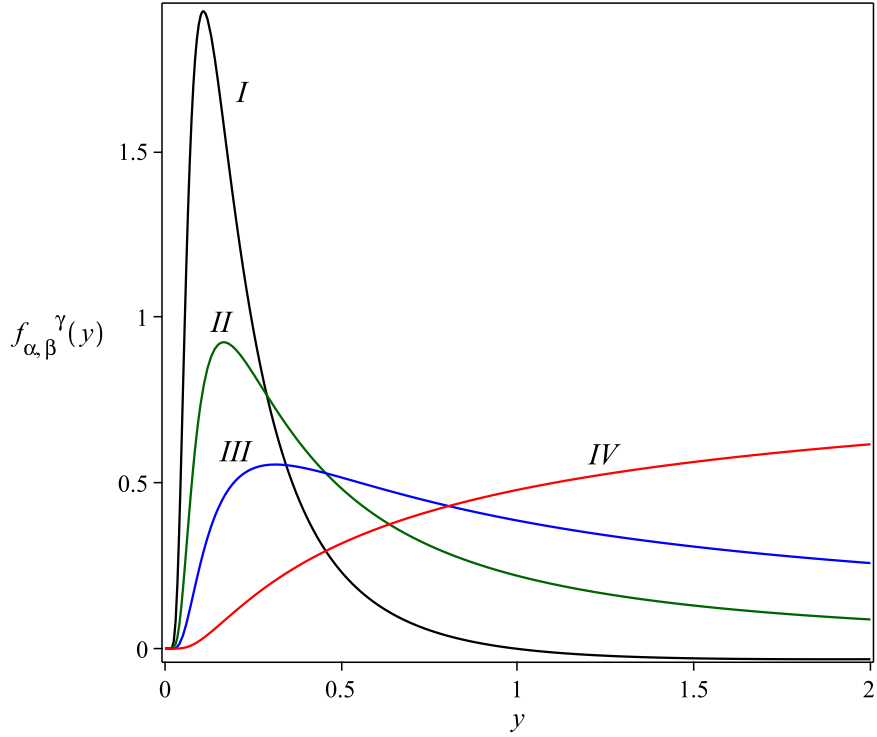}
\caption{\label{fig1} The plot of $f_{\alpha, \beta}^{\gamma}(y)$, $y > 0$, given by Eq. \eqref{8} with $\alpha = 1/2$ and $\gamma = 4/3$, for $\beta = 1/3$ (the "wrong" case $\alpha\gamma>\beta$, black line, no. I), $\beta = 2/3$ (the green line, no. II), $\beta = 1$ (the blue line, no. III), and $\beta = 5/3$ (the red line, no. IV).}
\end{center}
\end{figure}

Examples (a)-(c) illustrate our statement that the three parameter generalized Mittag-Leffler function $E_{\alpha,\beta}^{\gamma}(-x)$ is completely monotone for $x\ge 0$ under the conditions $0 < \alpha <1$, $\alpha\gamma \leq \beta$ and $\gamma, \beta > 0$ not bounded from above. Positivity check of $f_{\alpha, \beta}^{\gamma}(y)$ for $y \geq 0$ shows the importance of the condition $\alpha\gamma \leq \beta$ - if it is broken, e.g., by taking $\alpha = 1/2$, $\beta= 3/2$, and $\gamma=4$ (for which $\alpha\gamma > \beta$) then $f_{1/2, 3/2}^{\,4}(y) = (1 - 2 y)\, \exp[-1/(4y)]/(4 \sqrt{\pi} y^{5/2})$ which is obviously negative for $y > 1/2$. Consequently, once more because of the Bernstein theorem, the function $E_{\alpha,\beta}^{\gamma}(-x)$ can not be completely monotone for such choice of parameters. Fig. \ref{fig1} shows how the function $f_{\alpha, \beta}^{\gamma}(y)$ behaves for various $\beta$ with $\alpha$ and $\gamma$ kept fixed.

\section*{Acknowledgments}
K.G., A.H. and A.L. were supported by the Polish National Center for Science (NCN) research grant OPUS12 no. UMO-2016/23/B/ST3/01714. K. G. acknowledges support from the NCN Programme Miniatura 1, project no. 2017/01/X/ST3/00130, and from the program "Iuventus Plus 2015-2016" of the Polish Ministry of Science and Higher Education (MNiSW, Warsaw, Poland), project no IP2014 013073.


\end{document}